\documentstyle[aps,epsf,rotate,multicol]{revtex}

\begin{document}

\draft

\title{Multifractal Properties of the  Random Resistor Network}

\author{M. Barth\'el\'emy{\footnote{Permanent address: CEA-BIII,
Service de Physique de la Mati\`ere Condens\'ee, France.}}$^{1}$,
S.V. Buldyrev$^{1}$, S. Havlin$^{2}$ and H.E. Stanley$^{1}$}

\address{ $^1$ Center for Polymer Studies and Dept. of Physics, Boston
		University, Boston, MA 02215 \\ 
		$^2$ Minerva Center and
		Department of Physics, Bar-Ilan University, Ramat-Gan
		52900, Israel}

\maketitle

\begin{abstract}
We study the multifractal spectrum of the current in the two-dimensional
random resistor network at the percolation threshold. We consider two
ways of applying the voltage difference: (i) two parallel bars, and (ii)
two points. Our numerical results suggest that in the infinite system limit,
the probability distribution behaves for small $i$ as $P(i)\sim 1/i$
where $i$ is the current. As a consequence, the moments of $i$ of order
$q\le q_c=0$ do not exist and all current of value below the most
probable one have the fractal dimension of the backbone. The backbone
can thus be described in terms of only (i) blobs of fractal dimension
$d_B$ and (ii) high current carrying bonds of fractal dimension going
from $1/\nu$ to $d_B$.
\end{abstract}



\begin{multicols}{2}


The transport properties of the percolating cluster have been the
subject of numerous studies\cite{Stauffer92,Coniglio89}. A particularly
interesting system is the random resistor network (RRN), where the bonds
have a random conductance. The random resistor network serves as a
paradigm for many transport properties in heterogeneous systems as well
as being a simplified model for fracture\cite{Roux90}.

The first studies of the RRN were devoted to effective properties of the
network (conductivity, permittivity, etc.)\cite{Landauer78,Bergman92},
but for many practical applications---such as fracture, and dielectric
breakdown\cite{Roux90}---the central quantity is the probability
distribution $P(i)$ of currents $i$. For instance, in the random fuse
network, it is the maximum current corresponding the hottest or ``red''
bonds which will determine the macroscopic failure of the
system\cite{Roux90}.

The probability distribution $P(i)$ has many interesting features, one
of which is
multifractality\cite{ARC85,Rammal85a,Rammal85b,ARC86,ARC87,Fourcade87,Meir88,Nagatani87,Nagatani88}:
in order to describe $P(i)$, an infinite set of exponents is
needed. This idea of multifractality was initially proposed to treat
turbulence\cite{Mandelbrot74} and later applied successfully in many
different fields, ranging from model systems such as DLA\cite{Lee89}
to physiological data such as heartbeat\cite{Ivanov99}.

It was first believed\cite{ARC86,ARC87} that the low current part of
$P(i)$ and of the multifractal spectrum follow a log-normal law as it is
the case on hierarchical lattices. It is now clear\cite{Aharony93}, that
for small currents, the current probability distribution follows a power
law $P(i)\sim i^{b-1}$ where $b\ge 0$. For large currents, there is a
weak dependence on the system size $L$. This is in contrast with small
currents which are governed by very long paths, and therefore depend
more strongly on $L$. It was suggested\cite{Batrouni88,Aharony93} that
the exponent $b$ of the low-current part has a $1/\log L$ dependence,
where $L$ is the system size. The asymptotic value $b_{\infty}$ of the
exponent $b$ is of crucial importance. If $b_{\infty}$ is finite and
positive, then a low current evolves on a subset with a fractal
dimension depending on its value.  On the other hand, if $b_{\infty}$ is
zero, then the low current part of the multifractal spectrum is flat and the
entire backbone is contributing to low currents. It is thus important to
understand if the apparent subset structure with different fractal
dimensions is a finite-size effect.

This problem was adressed by Batrouni et al\cite{Batrouni88} who
conjectured a zero asymptotic slope and by Aharony et al\cite{Aharony93}
who proposed a finite asymptotic value. The maximum value of $L$ in the
literature is $128$\cite{Batrouni88}, so numerical estimates could not
lead to a definite conclusion. In this Letter, we present evidence that
the asymptotic slope is zero.


We first recall the basis of multifractality applied to the percolating
two-dimensional resistor network of linear size $L$. Let $n(i,L)$ be the
number of bonds carrying current $i$. By the steepest descent method,
the main contribution to $n(i,L)$ for large $L$ is given
by\cite{ARC85,ARC86,ARC87}
\begin{equation}
n(i,L)\sim L^{f(\alpha,L)}
\end{equation}
where $\alpha\equiv-\log i/\log L$. The multifractal spectrum
$f(\alpha,L)\equiv\log n/\log L$ can thus be interpreted as the fractal
dimension of the subset of bonds carrying the current $i$. The $q$-th
moment of the current is defined as $M_q\equiv\langle\sum i^q\rangle$,
where the sum is over all bonds carrying a non-zero current and
$\langle\cdot\rangle$ denotes an average over different disorder
configurations. These moments exists for $q>q_c$, and it can be easily
shown\cite{Aharony93} that the ``threshold'' is $q_c=-b$. The asymptotic
slope thus give the asymptotic value of the threshold $q_c$.

For the fixed current ensemble, one observes
that\cite{ARC85,ARC86,ARC87}$M_q\sim L^{\tau_q}$ for large $L$ and for $q>q_c$
and where $\tau_q$ is a universal exponent. In particular, $\tau_0=d_B$,
$\tau_2=t/\nu$, and $\tau_{\infty}=1/\nu$\cite{Coniglio82} where $d_B$
is the fractal dimension of the backbone, $t$ the conductivity exponent,
and $\nu$ is the correlation length exponent. If the behavior is
monofractal, then $\tau_q$ is a linear function of $q$, while in the
multifractal case, the exponents are not described by a simple linear
function of $q$. In the $L\rightarrow\infty$ limit, knowing $f(\alpha)$
is equivalent to knowing the infinite set of exponents $\tau_q$, as
$f(\alpha)$ is the Legendre transform of $\tau_q$\cite{Roux90}.

The low current part of $f(\alpha ,L)$ was found numerically to be a
power law of slope $b=b(L)$, where\cite{Batrouni88}
\begin{equation}
\label{bl}
b(L)=b_{\infty}+\frac{A}{\log L}+\varepsilon(L)
\end{equation}
and $\varepsilon(L)$ is a correction decreasing faster than $1/\log L$
when $L$ is increasing. This equation shows a strong finite-size effect
since $\log L$ grows very slowly, and two possibilities for $b_{\infty}$
were proposed, $b_{\infty}=0$\cite{Batrouni88} or
$b_{\infty}=1/4$\cite{Aharony93}.


We consider the two-dimensional random resistor network at criticality,
i.e. the fraction of conducting bonds $p$ is equal to its critical value
$p=p_c=1/2$. We first apply a voltage difference between two parallel
bars. We compute $f(\alpha ,L)$, for a fixed voltage difference, for
$L=50,\dots ,1000$, and average over $10^4$ configurations for each
$L$. We show our results in Fig.~1a. The slope is clearly decreasing
with $L$, confirming the strong finite size effects already
observed\cite{Batrouni88,Aharony93}.

Next, we consider a second type of configuration, which we call the
``two injection points'' case, in contrast with the usual ``parallel
bars'' case. We impose a voltage difference between two points $P$ and
$Q$ separated by a distance $r$, and we look for the backbone connecting
these two points. This situation was studied
in\cite{Lee99,Barthelemy99}, but here we keep only the backbones of size
$L$. In this way, we have large backbones connecting the two points $P$
and $Q$, and for $r\ll L$ we expect to have a large number of small
currents on bonds belonging to long loops. The multifractal spectrum is
then defined in the same way as for the parallel bars and we calculate
for different values of $L$ the slope of its small-current part. The
multifractal spectrum in this case is shown in Fig.~1b. We observe that
there is a large amount of small currents, and that the asymptotic limit
is reached faster in the two injection points case. We expect that the
low current distribution will be asymptotically the same as in the
parallel bar case, so the consistency between the two configurations will
support our results. However, for large currents there are some
distinct differences in the multifractal spectrum\cite{Noteoncurrent}.

Fig.~2a shows the slope $b$ versus $1/\log L$ according to Eq. (\ref{bl})
for both multifractal spectra. The extrapolation to $L=\infty$ is
consistent with $b_{\infty}=0$ in both cases. This result is consistent
with the behavior of the successive intercepts (Fig.~2b).

Another functional form of $b$ versus $L$ could lead to another value of
$b_{\infty}$. If we replace the abscissa of Fig.2(a) by $1/(\log
L)^{\kappa}$, then we find that the extrapolated value for $b_{\infty}$
depends on $\kappa$, ranging from $b_{\infty}\simeq 0.10$ for $\kappa=2$
to $b_{\infty}<0$ (which is impossible) for $\kappa=0.5$. It is
numerically difficult to distinguish between a $1/\log L$ and a $1/(\log
L)^2$ behavior, but the $1/\log L$ is the most commonly
used\cite{Batrouni88,Aharony93}.

In Fig.~2a, we observe higher order corrections to the behavior
$b(L)=b_{\infty}+A/\log L$. A better fit can be obtained by adding to
the linear form a small quadratic term $B/(\log L)^2$ (and eventually
even cubic and quartic terms). We find that we cannot do a quadratic fit
over the whole range of $1/\log L$, and indeed this leads to two
non-physical results: (a) For both geometries, the fits have {\it
negative} slopes at $1/\log L=0$, which is not physical since the larger
the system size, the larger the number of small currents, so the
behavior of $b$ should be monotonically decreasing with $L$. (b) A
second defect of these quadratic fits is that the obtained values for
the intercepts are {\it different} for the two geometries, which is
impossible.

The important assumption here is the behavior of the leading term. There
is no proof that the leading term of the expansion is $1/\log L$ rather
than $(1/\log L)^{\kappa}$ with $\kappa\neq 1$. However, the assumption
that the leading term of the expansion is $1/\log L$ with $b_{\infty}=0$
is consistent with our numerical data, and shows that the correction
$\varepsilon(L)$ decays faster than an inverse power of $\log L$ (see
Fig.~2c).

Finally, we note that the sequence of maximum values of $f(\alpha ,L)$
for the two injection points case plausibly extrapolates in the variable
$1/\log L$ as $L\rightarrow\infty$ to a value of $d_B$ close to the
known value $1.64$ (Fig.3).

Thus our results suggest the intriguing possibility that for
$L\rightarrow\infty$, the small current part of $f(\alpha ,L)$ is a
horizontal line at the value $d_B$, implying that in an infinite
system the fractal dimension of the subset contributing to small
current is $d_B$, independently of the value of $\alpha$. In this
sense, the small current probability distribution is apparently not
multifractal. The ``perfectly balanced'' bonds which carry zero
current have a fractal dimension equal to
$d_B$\cite{Batrouni88}. Since these bonds contribute to $f(\alpha ,L)$
for $\alpha\rightarrow\infty$, the fact that their fractal dimension
is $d_B$ supports our hypothesis that $b_{\infty}=0$. A related
conclusion is that $q_c=0$, or the negative moments of the current do
not exist in the infinite-size limit. In particular, it shows that the
first-passage time for a tracer particle travelling in a flow field in
a porous medium modelled by a percolation cluster diverges in an
infinite system.

Moreover, the result $b_{\infty}=0$ is supported by the following
argument. If $b_{\infty}$ were not zero, then the number of bonds carrying
a small current $i$ would be $n(i\to 0,L=\infty)\sim i^{b_{\infty}}$. This
behavior would indicate that the number of bonds carrying a small current $i$
approaches zero when $i\to 0$, which seems unlikely, since
on an infinite backbone, the number of loops is very large, and $n(i\to
0,L=\infty)$ should be nonzero. Hence $b_{\infty}=0$. This argument is
consistent with the fact that the total number of bonds carrying a
nonzero current, $\int_0 n(i,L)d(\log i)$, should diverge as $L\to\infty$.

For large values of the current, the multifractal features do not change
as $L$ increase, suggesting that in the infinite-size limit,
there are essentially two different type of subsets. The first comprises
the blobs of fractal dimension $d_B$, and the second set comprises links
carrying larger values of the current (red bonds), of fractal dimension
ranging from $d_{\text{red}}=1/\nu$ to $d_B$.

We thank L.A.N.~Amaral for valuable help, and J.S.~Andrade, A.~Chessa,
A.~Coniglio, N.V.~Dokholyan, P.~Gopikrishnan, P.R.~King, G.~Paul,
A.~Scala, and F.W.~Starr for useful discussions, two anonymous referees
for helpful suggestions, DGA and BP Amoco for financial support.



\begin{figure}
\narrowtext
\centerline{
\epsfysize=0.9\columnwidth{\rotate[r]{\epsfbox{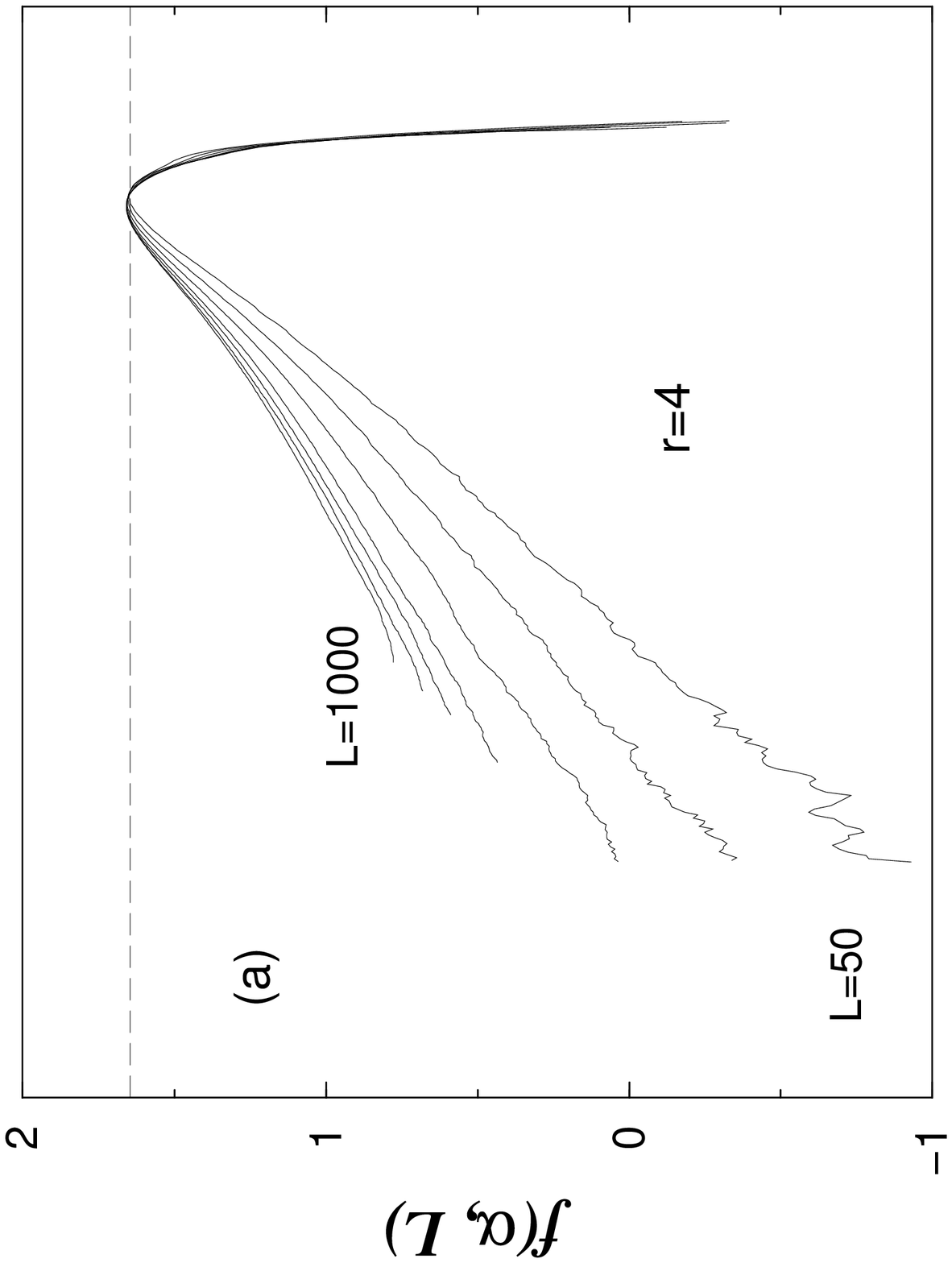}}}
}
\centerline{
\epsfysize=0.9\columnwidth{\rotate[r]{\epsfbox{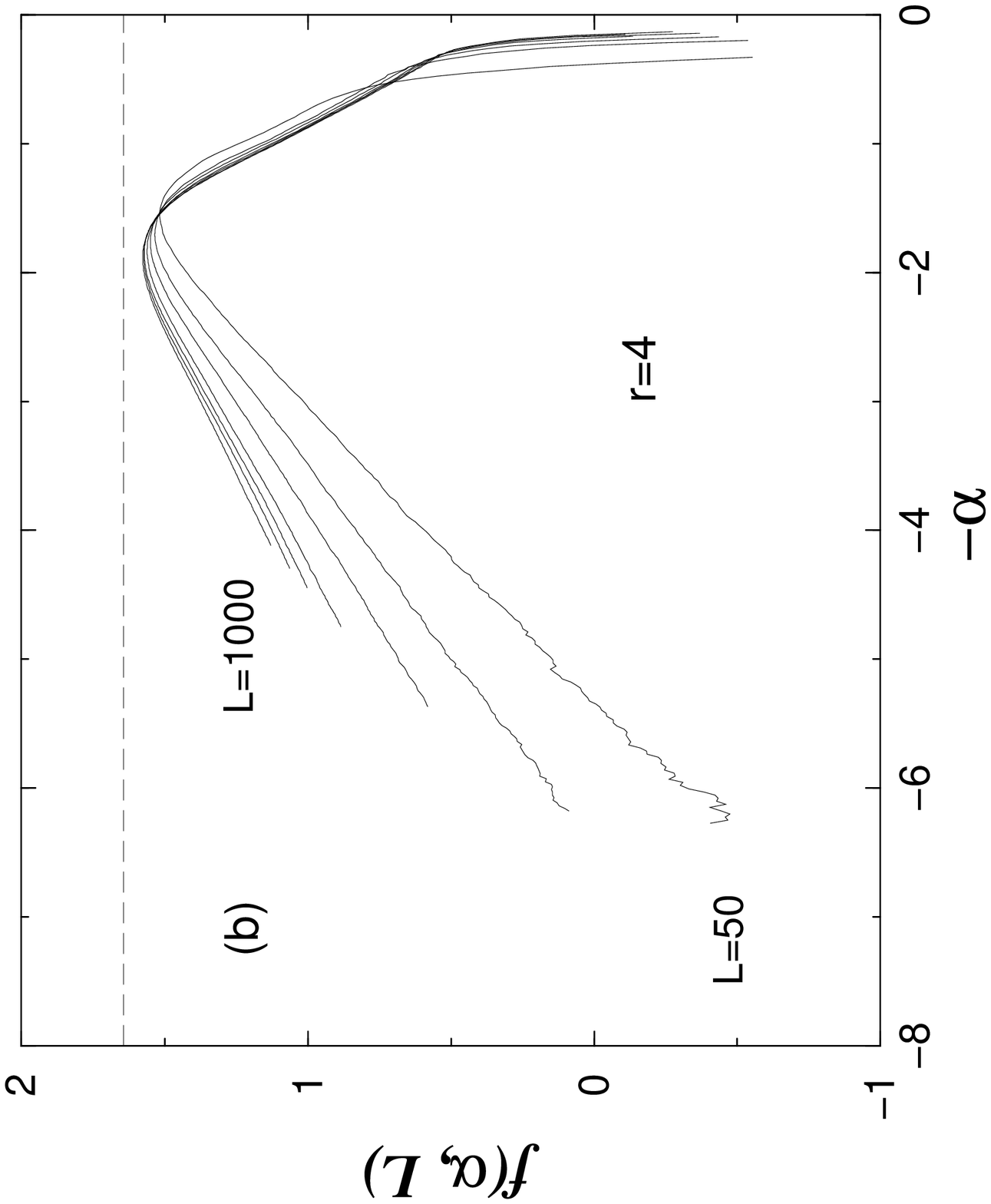}}}
}
\vspace*{0.0cm}
\caption{ Multifractal spectra for fixed voltage for (a) parallel bars
and (b) for two injection points separated by a distance $r=4$. We show
the results for $7$ different values of $L=50, 100, 200, 400, 600,
750,1000$ and averaged over $10^4$ configurations ($L$ increases from
bottom to the top). The horizontal dashed line is at $d_B\simeq 1.643$.}
\label{figure1}
\end{figure}

\vfill
\eject

\begin{figure}
\narrowtext
\centerline{
\epsfysize=0.9\columnwidth{\rotate[r]{\epsfbox{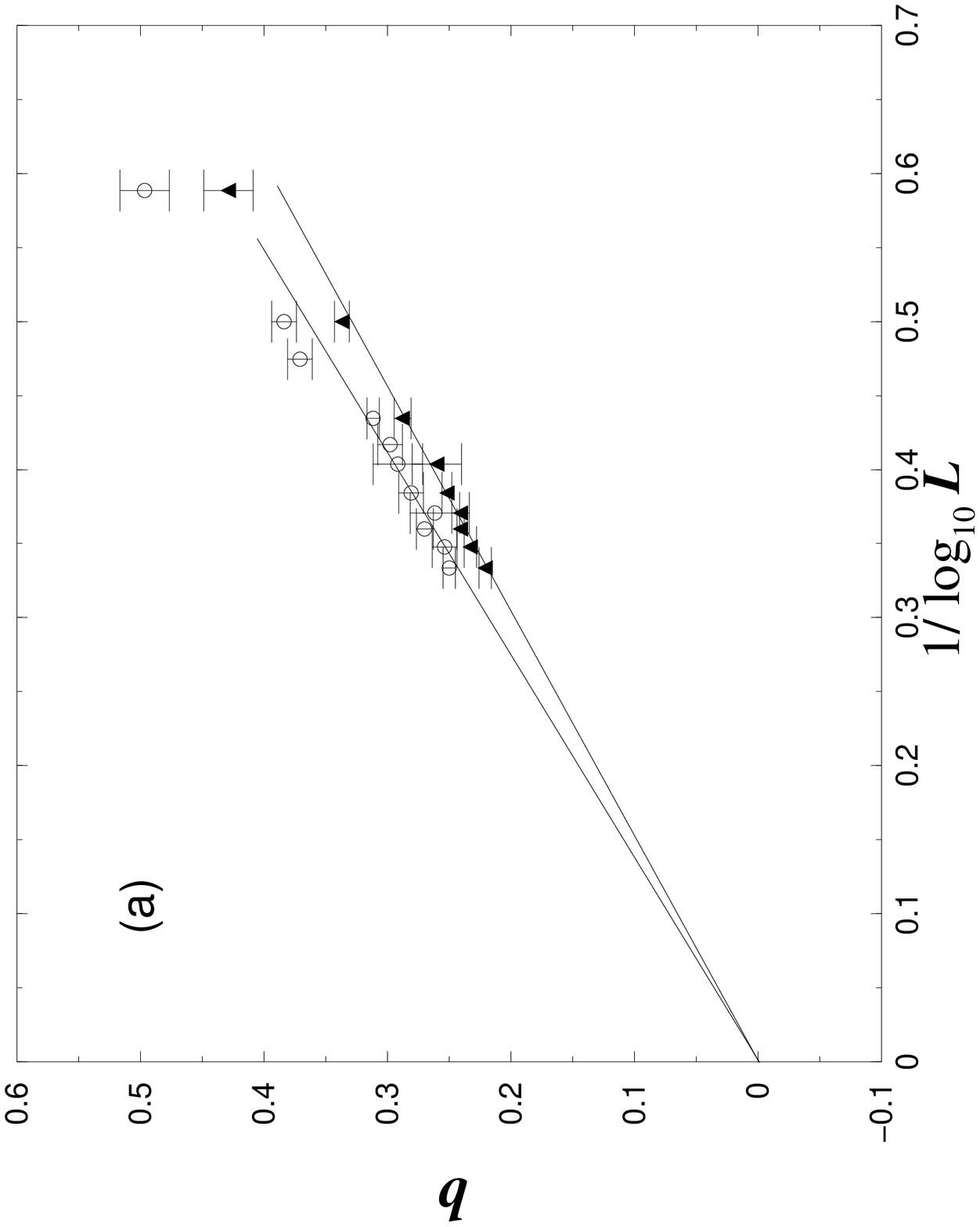}}}
}
\centerline{
\epsfysize=0.9\columnwidth{\rotate[r]{\epsfbox{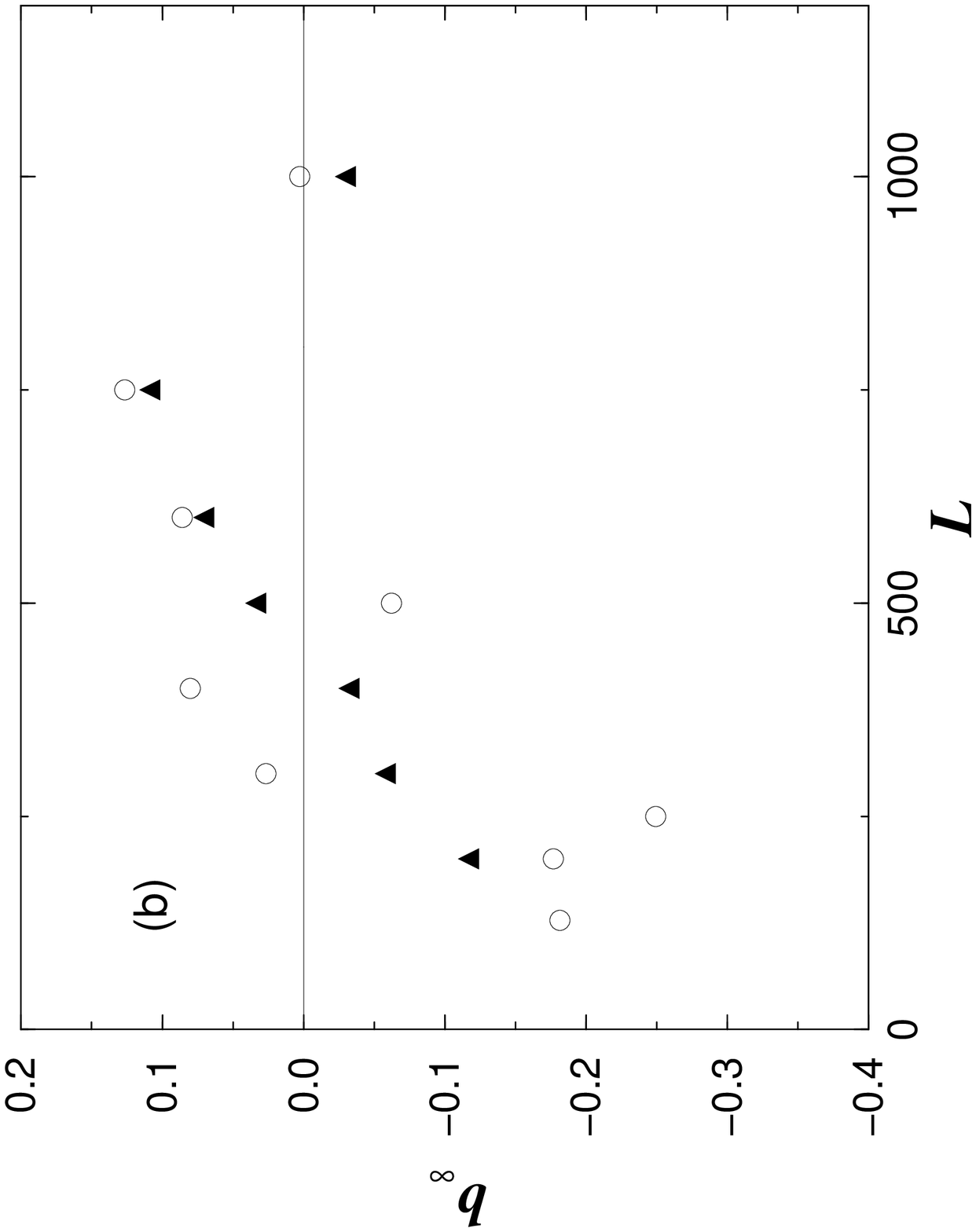}}}
}
\centerline{
\epsfysize=0.9\columnwidth{\rotate[r]{\epsfbox{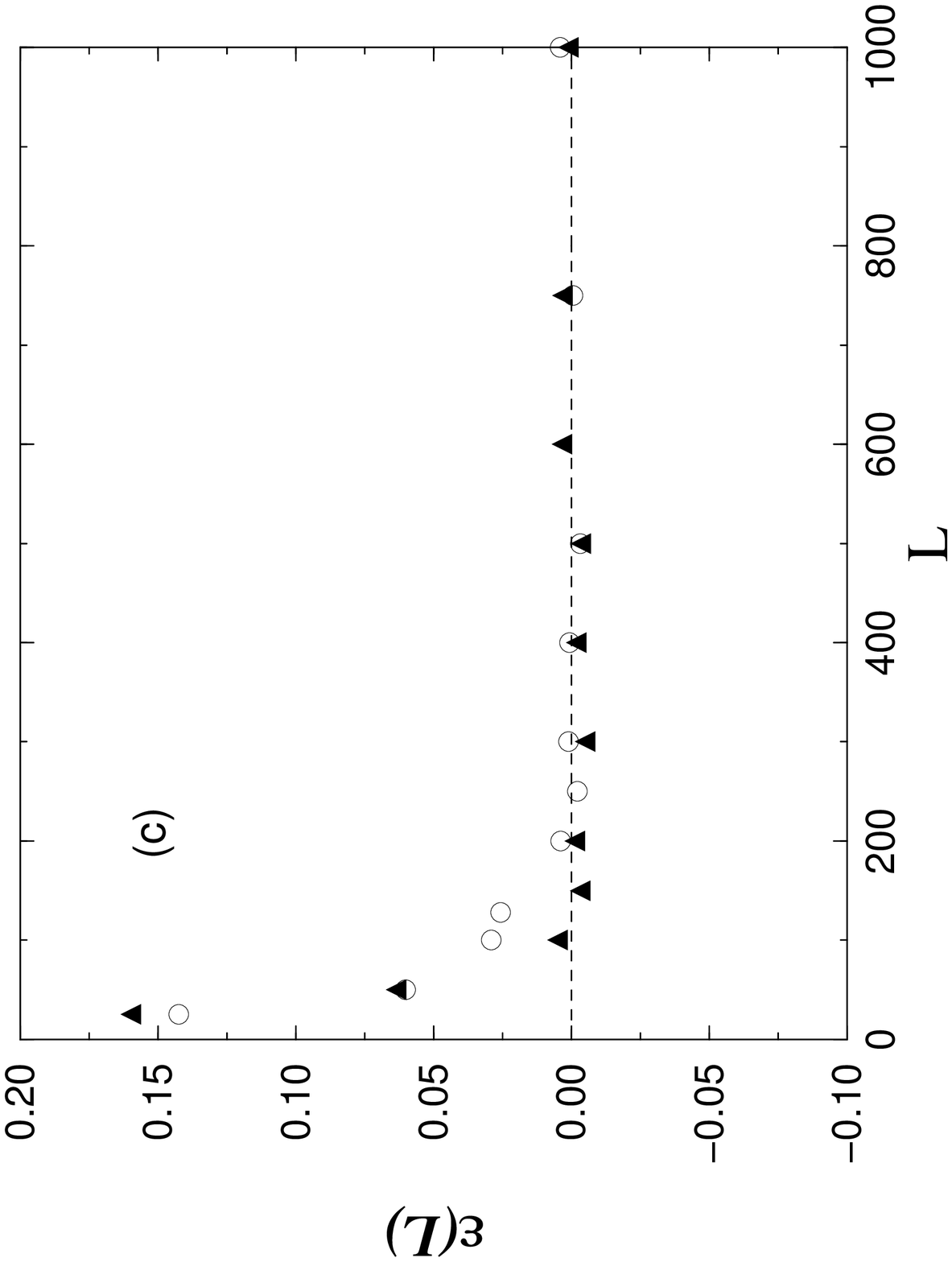}}}
}
\vspace*{0.0cm}
\caption{ (a) Slope $b$ versus $1/\log L$. The circles correspond to the
parallel bars case and the triangles to the two injection points
case. These values were obtained by fitting the small current parts of
Figs. 1a,b, roughly over the range $2.5<\alpha <5$. The extrapolation
shown as a guide to the eye is consistent with $b_{\infty}=0$. The error
bars where estimated by computing the local slopes and are going from
$0.02$ to $0.005$ as $L$ increases. (b) Successive intercepts computed
by using a least square fit over three successive points. The circles
correspond to the parallel bars case, the triangles to the two injection
points case. These plots are consistent with $b_{\infty}=0$. (c)
Correction $\varepsilon(L)$ for $L=25$ to $L=1000$ as given by
Eq.~(2). This plot shows the fast decay of the correction.}
\label{figure2}
\end{figure}

\begin{figure}
\narrowtext
\centerline{
\epsfysize=0.9\columnwidth{\rotate[r]{\epsfbox{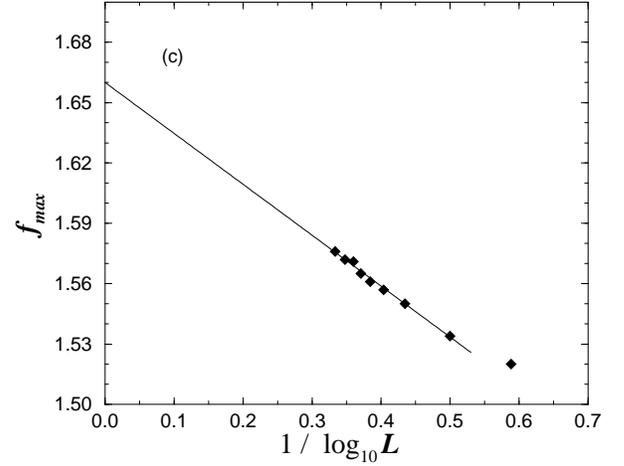}}}
}
\vspace*{0.0cm}
\caption{ The maximum of $f(\alpha ,L)$, $f_{max}(L)$, in the the two
injection points case vs. $1/\log L$. The least square fit shown gives
the extrapolated value at $L\rightarrow\infty$ of $d_B$ close to the
accepted value of $1.64$. Also, we found that a log-log plot of
$d_B-f_{max}(L)$ vs. $L$ is remarkably straight, with slope $\simeq
-0.20$.}
\label{figure3}
\end{figure}


\end{multicols}


\begin{references}
\bibitem{Noteoncurrent} In the ``two injection points'' configuration,
the large currents are controlled by the small loops between $P$ and
$Q$. The fluctuation of the length of the shortest path between $P$ and
$Q$ gives thus rise to the region of power law behavior for large currents we
observe in Fig.1b. For the constant current ensemble, the distribution
converges to $1/\nu$ for the largest current.
\end{references}
\end{document}